\documentclass[aps,pra,onecolumn,superscriptaddress,nofootinbibt,longbibliography]{revtex4-1}
\usepackage[utf8]{inputenc}
\usepackage[english]{babel}
\usepackage[T1]{fontenc}
\usepackage{amsmath}

\usepackage{tikz}
\usepackage{lipsum}

\usepackage{ulem}
\usepackage{amsmath}
\usepackage{amsfonts}
\usepackage{amssymb}
\usepackage{fancyhdr}
\usepackage{color}
\usepackage[utf8]{inputenc}
\usepackage{mathtools}
\usepackage{graphicx}
\usepackage{float}
\usepackage{physics}
\usepackage{bbold}
\usepackage{url} 
\usepackage{color}
\usepackage{cancel}
\usepackage{array} 
\usepackage{latexsym}
\usepackage{mathrsfs}
\usepackage{enumitem}
\DeclareGraphicsExtensions{.pdf,.png,.jpg}
\usepackage{array}
\usepackage{verbatim}
\usepackage{booktabs}
\usepackage{yfonts}
\usepackage[title]{appendix}
\usepackage[colorlinks=true, urlcolor=blue,citecolor=blue,linkcolor=blue]{hyperref}
\newcommand{\ave}[1]{\langle #1 \rangle}

\begin{document}

\title{Quantum Thermal Amplifiers with Engineered Dissipation} 

\author{Antonio Mandarino}
\affiliation{International Centre for Theory of Quantum Technologies (ICTQT), University of Gdansk, 80-308 Gdansk, Poland}

\begin{abstract}
A three-terminal device, able to control the heat currents flowing through it, is known as a quantum thermal transistor whenever it amplifies  two output currents as a response to the external source acting on its third terminal.  Several efforts have been proposed in the direction of addressing different engineering options of the configuration of the system. 
	Here, we adhere to the scheme in which such a {device} is implemented as a three-qubit system that interacts with three separate thermal baths. However, another interesting direction is how to engineer the thermal reservoirs to magnify the current amplification. Here, we derive a quantum dynamical equation for the evolution of the system to study the role of distinct dissipative thermal noises. We compare the amplification gain in different configurations and analyze {the role of the correlations in a system exhibiting the thermal transistor effect, via measures borrowed from the quantum information theory.}

\end{abstract}

\maketitle

\section{Introduction}
In nature, energy transport in condensed matter systems can be achieved mainly via two different mechanisms, namely electric and thermal conduction. However, despite their comparable relevance, they have been treated in a different way for many years. In~fact, the~  flow of electrons has boosted all the recent information and communication technologies, while heat production has always been seen as a detrimental effect. 
Recently, due to the terrific success of electronics, several contributions have opened a prolific field of investigation proposing devices that operate by exchanging heat instead of~electrons.

Heat can be understood as vibrations in the lattice structure of a solid. This collective behaviour at the microscopic level is described by the quantization of the modes of vibration, and~the resulting bosonic quasiparticles are known as phonons. 
Recently, the~engineering of the exchange of phonons paved the way to the exploitation of the heat flux and opened an intense field of research known as phononics~\cite{wang2008phononics, phononics}.
{In fact, at~the very basic level, the~thermal currents throughout an object in contact with to two or more thermostats are mediated by phonons.}

Since then, several different devices operating in an analogous way to the known electronic ones have been addressed in the literature, the~most notable are heat valves~\cite{heatvalve}, thermal rectifiers~\cite{diode1, diode2, diode3, diode4}, amplifiers~\cite{thermalTr,QTT1, QTT2, QTT3, QTT4} and thermal logic gates~\cite{ThermalGate}.

{A three-terminal device showing at the outputs an amplification of the currents
stays at the cornerstone of the modern development of electronic devices; for this reason, 
one of the devices that has attracted more attention, for~its possible applications, is the quantum thermal transistor. }

During the last few years, several possible implementations of this device have been proposed considering tree--qubit systems~\cite{QTT1, QTT2}, 
qubit--qutrit system~\cite{QTT3}, in~circuits of superconducting qubits~\cite{QTT4} and in a system with three-body interaction~\cite{liu2021common}.  
Eventually, also, networks of connected thermal transistors were proposed~\cite{QTTn1, QTTn2}. 
{The main trait of all these implementations is that they propose a three-terminal system to control the thermal energy exchanged by two of its terminals via an incoherent operation offered by the tuning of the temperature of its third terminal.} 

However, even if detailed studies about several implementations of the system have been proposed, less attention has been paid to the reservoirs' characteristics that lead to transistor-like behaviour. In~this paper, we tackle this second point and we address how the current amplification varies in function of two of the most significant phenomenological features of the thermostats: the temperatures and the noise spectra. Eventually, a~preliminary study on the correlation among the parties composing the system will allow us to grasp useful insight into the {non-equilibrium} steady state configuration of the~system. 

The paper is structured as follows: in Section~\ref{sec:Model} we describe the Hamiltonian of the system, while in Section~\ref{sec:DissDyn} we address its dissipative evolution. The~results regarding the heat amplification are presented in Section~\ref{sec:QTT}, and~in Section~\ref{sec:Corr} the role of  correlations arising from the dissipative dynamical evolution is studied. Finally, Section~\ref{sec:Outro} closes the paper with the final remarks and avenues for future~research. 

We remark from now that throughout the article, we will use a system of natural units 
fixing $\hbar = c = K_B = 1,$ 
where $\hbar $ is the Planck’s constant, $c$ is the speed of light in vacuum 
and $K_B$  is the Boltzmann’s~constant.

\section{The~Model}
\label{sec:Model}
\begin{figure}[H]
\includegraphics[width= 0.70 \columnwidth]{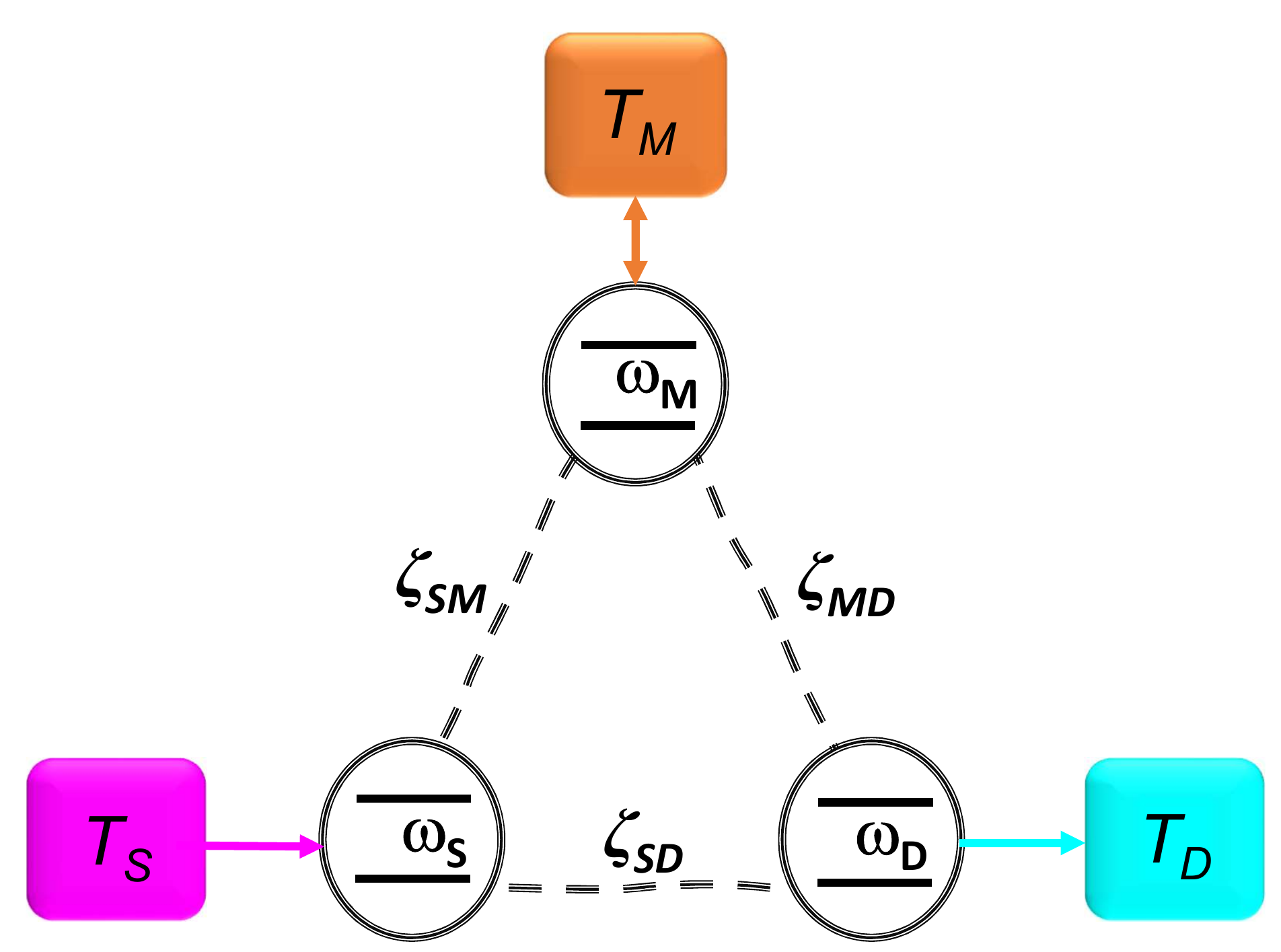}
\caption{\label{fig:1} Representation of the dissipative model considered in the paper.
A system of three coupled qubits as in Equation~\eqref{eq:hsys}, connected to independent thermal reservoirs as described in  Equation~\eqref{eq:Bathhamiltonians} 
via the interaction in Equation~\eqref{eq:Vint}. }
\end{figure}

The effects of different forms of dissipation on the
transport of thermal excitation in a quantum system are described considering first of all 
a generic model Hamiltonian:
\begin{equation}
    H= H_S + H_R + V_{SR}. 
\end{equation}
The three terms in the previous equation correspond to the system Hamiltonian ($H_S$), the~Hamiltonian of the thermal reservoirs ($H_R$) and their interaction $(V_{SR})$. The~first term is the free Hamiltonian of three-qubits. Each of them occupies a vertex of a triangular graph. For~ease of notation and analogy with the electronic terminology, we label them as $S$, $M$ and $D$, (standing for source, modulator and drain, respectively) as depicted in Figure~\ref{fig:1},  and~it explicitly reads:
\begin{equation}
\label{eq:hsys}
H_{S}= \frac{1}{2} \sum_{k=S,M,D} \omega_k \sigma^z_k + \sum_{n<k}  \zeta_{kn} \sigma^y_k \sigma^y_n, 
\end{equation}
where {$\sigma^\alpha_k$} is the $\alpha-$th Pauli matrix acting on the $k-$th qubit. In~particular, we chose as the magnetization easy axis the $z$ axis and defined the eigenstates of $\sigma^z \ket{j} = (-1)^j \ket{j}$ {as} a computational basis. 
{They constitute the smallest realization of a fully connected Ising model in a transverse field, also known as Lipkin model. It can be implemented in nanostructurated systems such as quantum dots or single-molecule magnets and has found application for several quantum technology purposes~\cite{LMG14, appLMG1, appLMG2, QTT2}. }

It is worth noting that the considered configuration in Figure~\ref{fig:1} has been widely studied also for quantum thermodynamics purposes. In~fact, 
a system composed by three qubits in thermal contact with the same number of baths has been proposed as the building block of a quantum absorption refrigerator
\cite{refr1,refr2,refr3,refr4,refr5}. The~purpose of this engine is to cool one of the three qubits. 
This is reached imposing that the intra-qubit interaction is weak enough to assume that each qubit dissipates only into the bath directly connected to itself. 
However, despite a similar configuration, a~main difference appears concerning the local nature of the dissipation. We will see in the following that the mechanism leading to a quantum thermal transistor is a collective one, i.e.,~the compound system as a whole interacts and exchanges energy with the three thermal environments.

The three reservoirs are assumed to be separate to avoid cross-talk dissipation~\cite{galve2017} and are described by
\begin{eqnarray}
\label{eq:Bathhamiltonians}
H_B &=& \sum_{k=S,M,D} h_k, \, \, \text{with} \, \,  h_k= \sum_{p} \varepsilon_p a_p^{k\,\dagger}a_p^{k}, 
\end{eqnarray}
while the system-reservoir interaction is linear and reads
\begin{eqnarray}
\label{eq:Vint}
V_{SR} &=&  \sum_{k=S,M,D} S_k \otimes R_k ,
\end{eqnarray}
where {$a^k_{p}$  ($a_p^{k\,\dagger}$)} are the annihilation (creation) operators of the reservoir $k.$
{We remark that each qubit is \emph{directly} coupled only to the corresponding $k-$th bosonic reservoir.}

\section{Non-Equilibrium~Dynamics}
\label{sec:DissDyn}
To describe the reduced dynamics of the system, we assume that the interaction between the system and the reservoir is weak 
and in a regime in which the Davies derivation of the Markovian master equation (MME) holds~\cite{davies1974markovian, alicki2007quantum}. 
Therefore, it has the following structure
\begin{equation} \label{eq:master}
\frac{\partial \rho}{\partial t}=-i[H_S+H_{LS},\rho] +\sum_{k=1}^3 \mathcal{D}_k[\rho],
\end{equation}
where $H_{LS}=\sum_{k,\,\omega}  f(\omega) S_k^{\dagger}(\omega) S_k(\omega) $
is the Lamb shift Hamiltonian, responsible for a shift in the system's frequency due its interaction with the reservoirs.  
The system operators $S_k(\omega)=\sum_{\omega=E_i-E_j} \ketbra{E_j}{E_j}S_k\ketbra{E_i}{E_i}$, 
are frequency-dependent and the sum is extended over all the eigenvalues of the system, $E_i,$ corresponding to the eigenvectors 
 $ \ket{E_i}$ with $i=  1, ..., 8 $, such that the difference has a fixed value of frequency $\omega$. 
 The following commutation relations hold: $[H_S, S(\omega)] = - \omega S(\omega)$ and $[H_S, S^\dagger(\omega)] =  \omega S^\dagger(\omega).$
  
The Gorini--Kossakowski--Sudarshan--Lindblad (GKSL) operators are given by~\cite{GKSL1, GKSL2}
\begin{eqnarray}
\mathcal{D}_k[\rho]=\frac12 \sum_{\omega>0} G_k(\omega) \Bigl( \Bigl[ S_k(\omega), \rho S_k^\dagger(\omega)\Bigr] + \Bigl[ S_k(\omega) \rho, S_k^\dagger(\omega)\Bigr] + \\ \nonumber
e^{-\omega/T_k} \Bigl[ S_k^\dagger(\omega), \rho S_k(\omega)\Bigr] + \Bigl[ S_k^\dagger(\omega) \rho, S_k(\omega)\Bigr] \Bigr). 
\label{eq:lindblad}
\end{eqnarray}
This is the general form for a Markovian master equation and, assuming that the state of each reservoir is $\nu_k$, their influence on the system relaxation processes is described by the power spectrum defined as:
\begin{eqnarray}
G_k(\omega)=\int_{- \infty}^{+\infty}e^{i \omega t} \Tr(R_k(t)R_k \nu_k) dt, 
\end{eqnarray}
where $R_k(t)$ are the reservoir operators entering the interaction Hamiltonian in the interaction picture. 
In deriving the equation in \eqref{eq:lindblad}, we have assumed that all  three baths fulfil the Kubo--Martin--Schwinger condition such that $G_k(-\omega) = e^{-\omega/T_k} G_k(\omega)$. Therefore, it is easy to identify the physical meaning of both terms  proportional to $G_k(\omega)$. The~first one characterizes the dissipation via emission of quanta of frequency $\omega$ into the $k-$th bath, while the second term corresponds to the absorption of quanta by the~system.

Each reservoir is in 
a Gibbs thermal state $\nu(T_k) = \frac{ e^{- h_k/ T_k}}{\Tr e^{- h_k/ T_k}}$, 
and we assume a system--reservoir interaction of the Caldeira--Leggett type~\cite{Rev_Leggett}, namely
\begin{eqnarray}
V_{SR}= \sum_{k=S,M,D} v_k, \, \, \text{with} \, \,  v_k= \sigma^y_k  \otimes \sum_{p} c_p^k  (a_p^{k\,\dagger} + a_p^{k}).
\end{eqnarray}
The coefficients $c_p$  account for the coupling strengths of the system 
to each mode of energy $\varepsilon_p$ in the~bath. 

Moreover, the~coefficients $G_k(\omega)$ can be expressed as a product of two terms $G_k(\omega)=J_k(\omega) (n_k(\omega) + 1),$ where 
$n_k(\omega)=\left[\mathrm{exp} \left(\omega/T_k\right)-1\right]^{-1}$ 
is the mean number of phonons, and~ $J_k (\omega)$ is the spectral density of the $k-$th reservoir. 
{The latter one gives information about the relevance of the noise at a given frequency $\omega_p$ and is determined by }
\begin{equation}
J_k (\omega) = \pi \sum_{p} \frac{|c^k_p|^2}{\omega_p} \delta(\omega- \omega_p). 
\end{equation}
In accordance with the notation of the  Caldeira and Leggett model~\cite{caldeira1981,caldeira1983}, we choose for the bosinic environments a spectral density of the following form
\begin{equation}
\label{eq:sd}
    J_k (\omega) = \lambda_k \omega_c \left( \frac{\omega}{\omega_c} \right)^s e^{-\frac{\omega}{\omega_c}}.
\end{equation}
The coefficient $\lambda_k$ determines the overall strength of the qubit--reservoir coupling and $\omega_c$ is a cutoff frequency that depends on the physical realization of the thermal reservoir. Throughout the paper we will assume $\omega_c = 10 \max_i \{E_i\}$, such that all the conditions on the different time scales needed to derive the MME are fulfilled~\cite{davies1974markovian}. 

This type of spectral density has been widely used to study the transport in {non-equilibrium} quantum systems such as quantum dots, nanotubes and molecular systems~\cite{SD1}. 
Its dependence upon the parameter $s$ allows us to identify three mean types of dissipation mechanism. The~sub-Ohmic case is for $s < 1$, the~Ohmic one is for $s=1$ and the super-Ohmic one for $s > 1$ \cite{Weiss}. 

\subsection*{Definition of Heat~Currents}

{Following the standard approach, we introduce the currents for time-independent system Hamiltonian in quantum thermodynamics (see, e.g.,~\cite{kosloff} for a detailed discussion). A~heat current flowing through a system in contact with multiple thermostats is defined as the time derivative of the system mean energy, namely }
\begin{eqnarray}
\label{eq:curr}
\mathcal{I}=\frac{\partial \ave{H_S}}{\partial t} = \Tr{ H_s \frac{\partial \rho}{\partial t}}. 
\end{eqnarray} 
For all our purposes we will consider the currents that are flowing through the system when it is in a Non-Equilibrium Steady State  (NESS), 
such that  $\frac{\partial \rho_{NESS}}{\partial t}= 0$. In~this configuration the preceding equation in \eqref{eq:curr} reduces to
\begin{eqnarray}
\label{eq:firstp}
0= \sum_{k=S,M,D}\mathcal{I}_k = \sum_{k=S,M,D} \Tr{ H_s \mathcal{D}_k[\rho_{NESS}] }.
\end{eqnarray} 
In the last equality we have used the property of the stationary solution of the GKSL equation in \eqref{eq:lindblad}, such that the NESS density matrix commutes with the system~Hamiltonian. 

{The chosen geometry depicted in Figure~\ref{fig:1} allow us to identify three different currents. For~shorthand of notation  we will refer to them as the \emph{source}, the~\emph{modulator} and the \emph{drain} current, and~label them as $\mathcal{I}_S$, $\mathcal{I}_M$ and $\mathcal{I}_D$.} They refer to the heat exchanged by the three-qubit system with the homologous reservoirs, respectively. 
For the sake of clarity, we remark here that we are in a regime of global dissipation, 
i.e., it is the whole system exchanging energy with the thermostats and not only the single qubit directly coupled with it~\cite{rivas2019}.

\section{Quantum Thermal~Transistor}
\label{sec:QTT}

In electronics, one of the main features of the transistor consists 
in the amplification of the currents at the source and the drain having an almost null current at the modulator. 
{The equivalence between electronic and thermal transistor is established when the fermionic leads are substituted by bosonic thermostats (described by a Gibbs state with null chemical potential). }
Temperatures and heat currents will play the part of the voltages and electronic currents, respectively. 
In particular, in~the thermal equivalent of an electronic transistor, we use as control parameter the temperature $T_M$ of the modulator thermostat, that will assume the role of the gate voltage. 
Using the tools developed in the preceding sections, we discuss how the temperature difference between the source and the drain, and~ 
different types of spectral density affect the amplification of the heat~currents.

\subsection{Amplification of Heat~Currents}
\begin{figure}[H]
\includegraphics[width= 0.85\columnwidth]{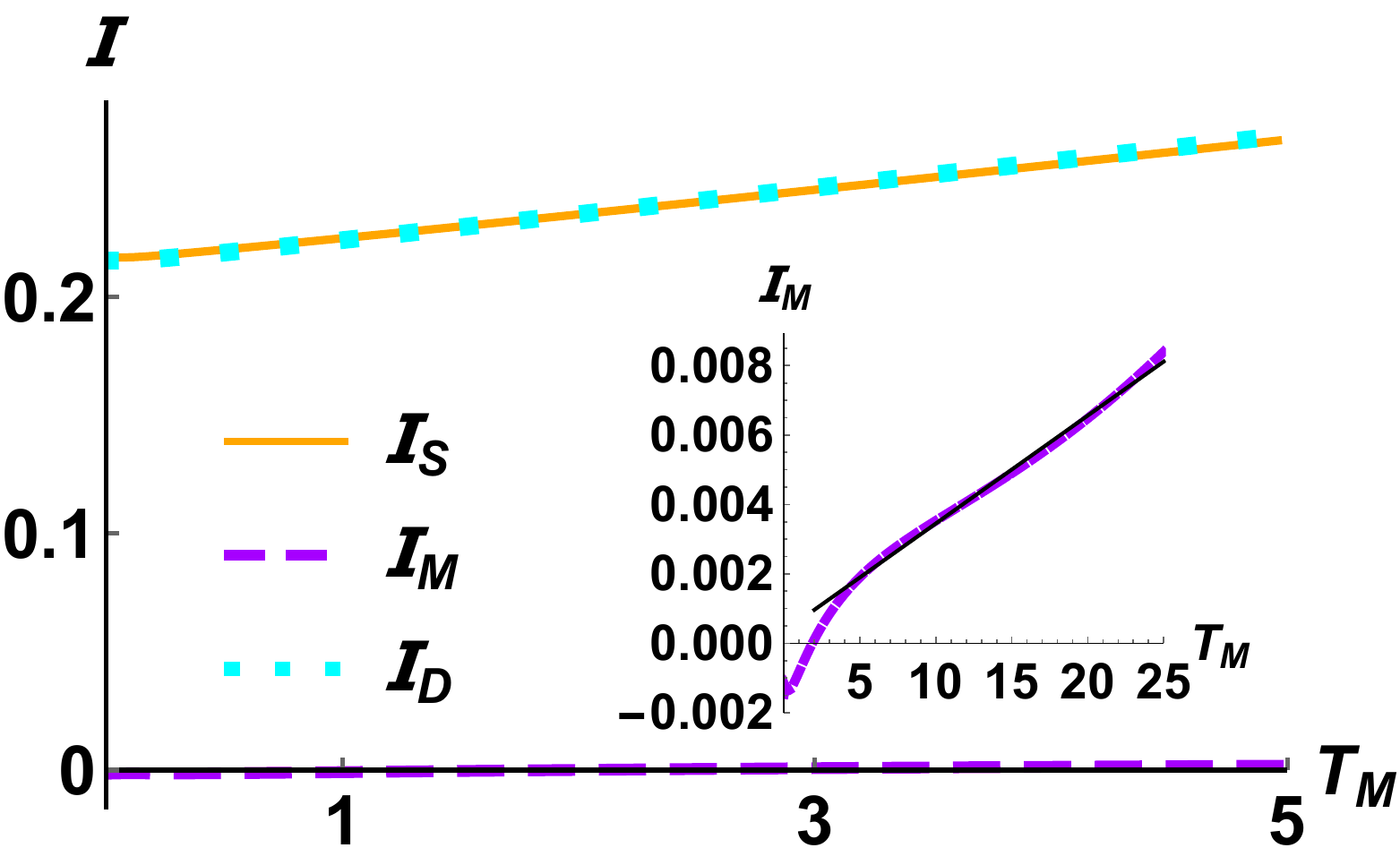}
\caption{\label{fig:curr} {The} 
 three thermal currents defined in Equation~\eqref{eq:curr} 
exchanged by the system with the three Ohmic reservoirs ($s=1$) as a function of the modulator temperature $T_M$. Upon~fixing the frequency of the source qubit as reference, namely $\omega_S= \omega$, 
the parameters are $\omega = 10 \omega_M=3\omega_D= \zeta_{SM}=6 \zeta_{MD}=\zeta_{SD}.$ The source and drain temperature are set to $T_S= 10 \omega$ and $T_D= 0.01 \omega$, respectively. 
The coupling strength of the system with the three reservoirs are $10^6\lambda_S=10^6\lambda_M=10^4\lambda_D= \omega$. Note that to highlight that the energy is conserved we plot $|\mathcal{I}_D|$. 
In the inset the current $I_M$ exchanged by the modulator reservoir with the system. The~black solid line is a linear fit $\mathcal{I}_M = m T_M + q$, with~$m \simeq 3.1 \times 10^{-4}$  and $q \simeq 3.4 \times 10^{-4}$.}
\end{figure}

{The first step towards the assessment of a thermal transistor is done by looking at the behaviour of the currents that that three-qubit system exchange with the reservoirs. In~analogy with the current-voltage characteristic curve for electronic components, in~Figure~\ref{fig:curr} we plot a paradigmatic example of the current--temperature curve for a system showing amplifications of currents at two terminals.}
The parameter we have chosen as control is the temperature of the collector, $T_M,$ and  
without loss of generality we imposed $T_S > T_D.$  
{It is easy to see how the currents of the configuration considered in Figure~\ref{fig:curr} exhibit the standard behaviour of those observed in a transistor: the source and the drain currents are amplified, while {the collector current remains} almost constant in the entire interval of temperature.} 
We report here $|\mathcal{I}_D|$, to~make it more visible that it is almost the exact opposite of $\mathcal{I}_S$  signaling a quasi-null heat flow between the three qubits and the modulator~thermostat.  

{In the inset of Figure~\ref{fig:curr}, we have reported the behaviour of the modulator current, in~the interval $0 \leq T_M \leq 5.$
It is easy to see that for $0 \leq T_M \leq 5$ where the source and drain currents are amplified, $|\mathcal{I}_M| \approx 0.01 \mathcal{I}$. On~the contrary, for~values of $T_M$ beyond such interval we observe a linear grow of the modulator current. 
In fact, when the temperature of the modulator {reservoir} is comparable or higher than the highest one (in the case considered $T_S$) 
it starts to inject heat into the system. In~other words, we can say that it stops to work as a buffer between the hot and the cold reservoir.} 

\subsection{Amplification~Factor}

We examine here the currents exchanged by the system with the three thermal environments 
depending on a suitable engineering of their characteristics, namely temperature $T_k$ and spectral density $J_k(\omega)$. 
Each configuration will be identified by the functional type of the considered spectral density (subOhmic $s=0.5$, Ohmic $s=1$ and superOhmic $s=1.5$) and by the temperature difference between the {source and drain reservoirs.} 

We have seen, in~Figure~\ref{fig:curr}, that  a variation of the gate temperature $T_M$ produces a significant variation of 
the two lateral currents in contrast with a significantly smaller value of $\mathcal{I}_M$.
However, observing the behaviour of the heat currents gives only a qualitatively assessment of the 
presence of an effect comparable to the amplification produced by a transistor.
To have a quantitative benchmark of the amplification as a function of the control temperature $T_M$
it is suitable to introduce the amplification factor:
\begin{equation}
\label{alpha}
\beta =\left | \frac{ \partial \mathcal{I}_S}{\partial \mathcal{I}_M} \right | = \left | \frac{ \partial \mathcal{I}_S}{\partial T_M} \left( \frac{ \partial \mathcal{I}_{M}}{\partial T_M} \right)^{-1} \right |
\end{equation}
In principle, one can define also the factor comparing the change {in the drain current over the modulator one}.
Anyway, given Equation~\eqref{eq:firstp}, that is a reformulation of the first principle of the thermodynamics in terms of currents, one can show that the relation $ \beta_S + \beta_{D} = -1$ holds. 

For our aim, the~interesting interval of temperature are those for which $|\beta|\gg 1.$ 
In fact, a~high value of this parameter signals a strong amplification of the currents 
{at the source and the drain compared to the one the system exchanges with the reservoir acting as modulator}.

\begin{figure}[H] 
\includegraphics[width= 0.50 \columnwidth]{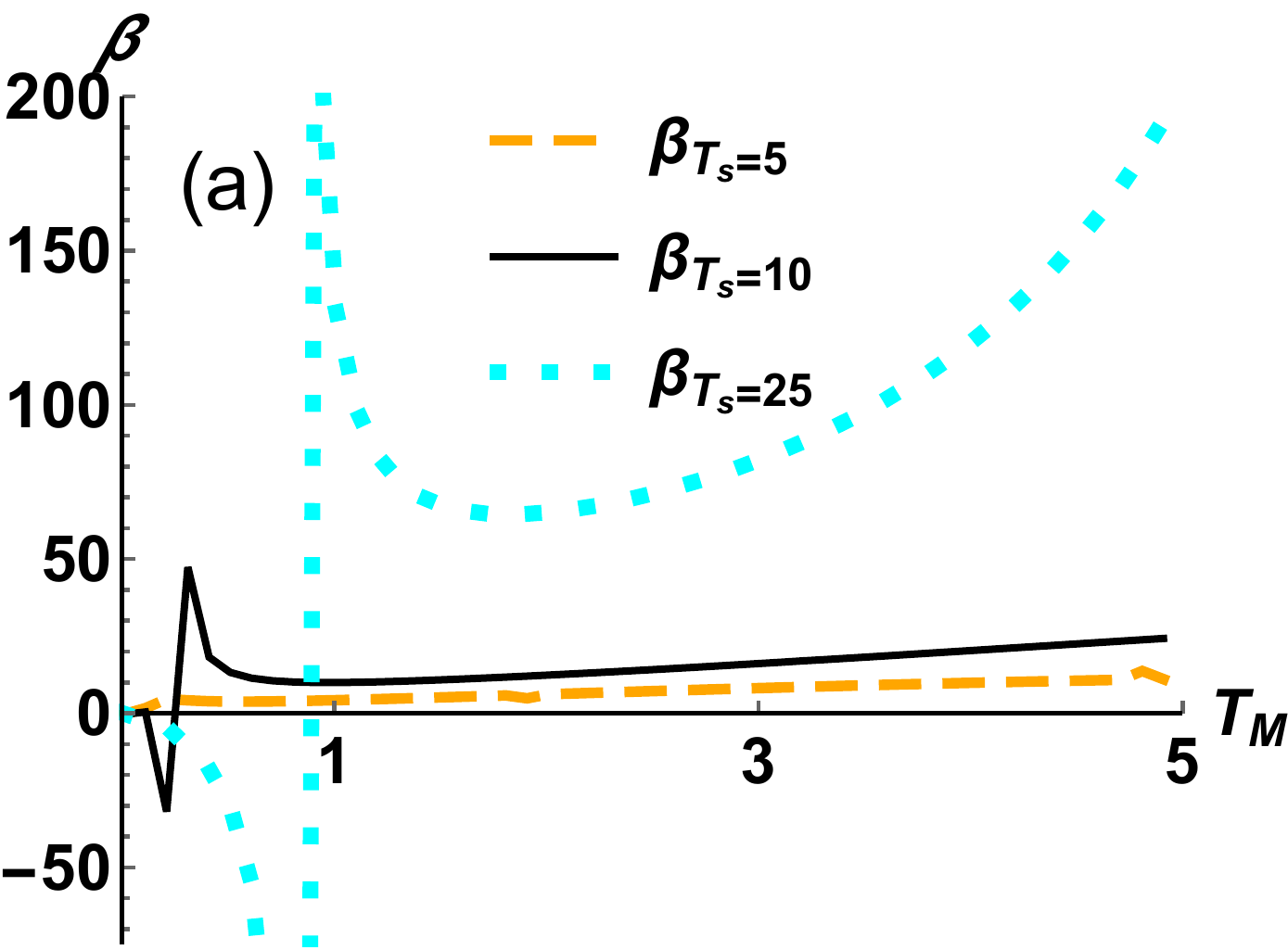}
\includegraphics[width= 0.50 \columnwidth]{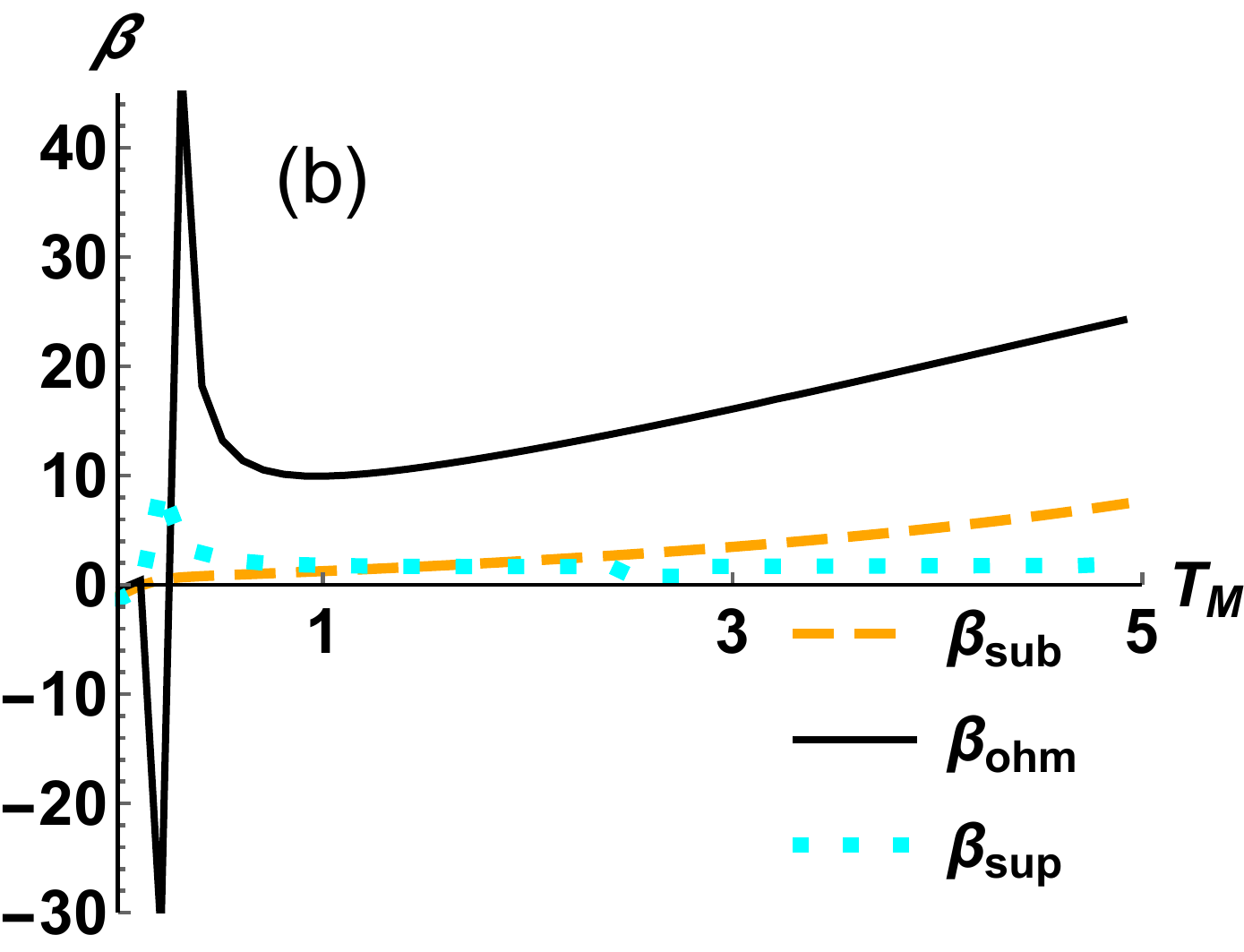}
\caption{\label{fig:amplif} We plot the amplification factors for different bath configurations, assuming as a reference the value giving the heat currents in Figure~\ref{fig:curr}.
(\textbf{a}) The configurations are given by the temperature difference between the hot and the cold thermostat. We fix $T_D$ and set $T_S=5 \omega$ (orange dashed line) and $T_S=25 \omega$ (cyan dotted line).
(\textbf{b}) The configurations are given by a different spectral density for the reservoirs. We assume for all the three baths a subOhmic $s= 0.5$ (orange dashed line) and superOhmic $s=1.5$ (cyan dotted line).
}
\end{figure}

We report in Figure~\ref{fig:amplif} the amplification factors for different reservoirs' engineering.  
In particular, we focus on the temperature gradient between the hottest and the coldest reservoir ($T_S - T_D$) and on the power noise of the the reservoirs via three paradigmatic cases of spectral densities as in Equation \eqref{eq:sd}. 
{For all the considered settings the operating regime of the QTT is given by the interval }in which the temperature of the gate reservoir varies, i.e.,~$T_M \in [0, 5]$.

In Figure~\ref{fig:amplif}a we observe that increasing the temperature gap, $T_S - T_D$, leads to a magnification of the thermal transistor effect.
This is achieved without performing any operation on the system. In~contrast, it is apparent from Figure~\ref{fig:amplif}b how a change in the dissipation model does not contribute to a better performance when building a QTT. 
In fact, a~sudden transition appears from subOhmic to superOhmic regimes in correspondence to a power noise with $s=1$.

\section{Insights into the Transistor Effect via Entropic Measures of~Correlations}
\label{sec:Corr}

In this section we study the behaviour of the correlations present in the three-qubit mixed state $\rho_{NESS}$ 
as a function of the modulator temperature $T_M$ for the different configurations of the reservoirs, as~discussed in the previous sections. 
Introducing the von Neumann entropy $\mathcal{S}$ of a quantum state $\rho$  (the quantum analogue of the Shannon entropy):
\begin{equation}
   \mathcal{S}(\rho) = - \Tr{\rho \log \rho } 
\end{equation}
we can consider the two and three particle mutual information~\cite{mutual_arul, rota2016tripartite, Rangamani_2015} as quantifier of the total correlations in the state. They are respectively defined as follows:
\begin{eqnarray}
\label{eq:mutualinfo2}
\mathcal{M}_2(\rho_{AB}) = \mathcal{S}(\rho_A)+\mathcal{S}\rho_B) - \mathcal{S}(\rho_{AB}),
\end{eqnarray}
and
\begin{eqnarray}
\label{eq:mutualinfo3}
\mathcal{M}_3(\rho_{ABC}) = \mathcal{S}(\rho_{ABC})+ \mathcal{S}(\rho_A)+ \mathcal{S}(\rho_B) + \mathcal{S}(\rho_C)  - \mathcal{S}(\rho_{AB}) - \mathcal{S}(\rho_{AC}) - \mathcal{S}(\rho_{BC})  
\end{eqnarray}
where the marginal states $\rho_X = \Tr_{k \neq X} \rho_{ABC} $ are the partial trace over the not considered qubit(s).  

\begin{figure}[H] 
\includegraphics[width= 0.7 \columnwidth]{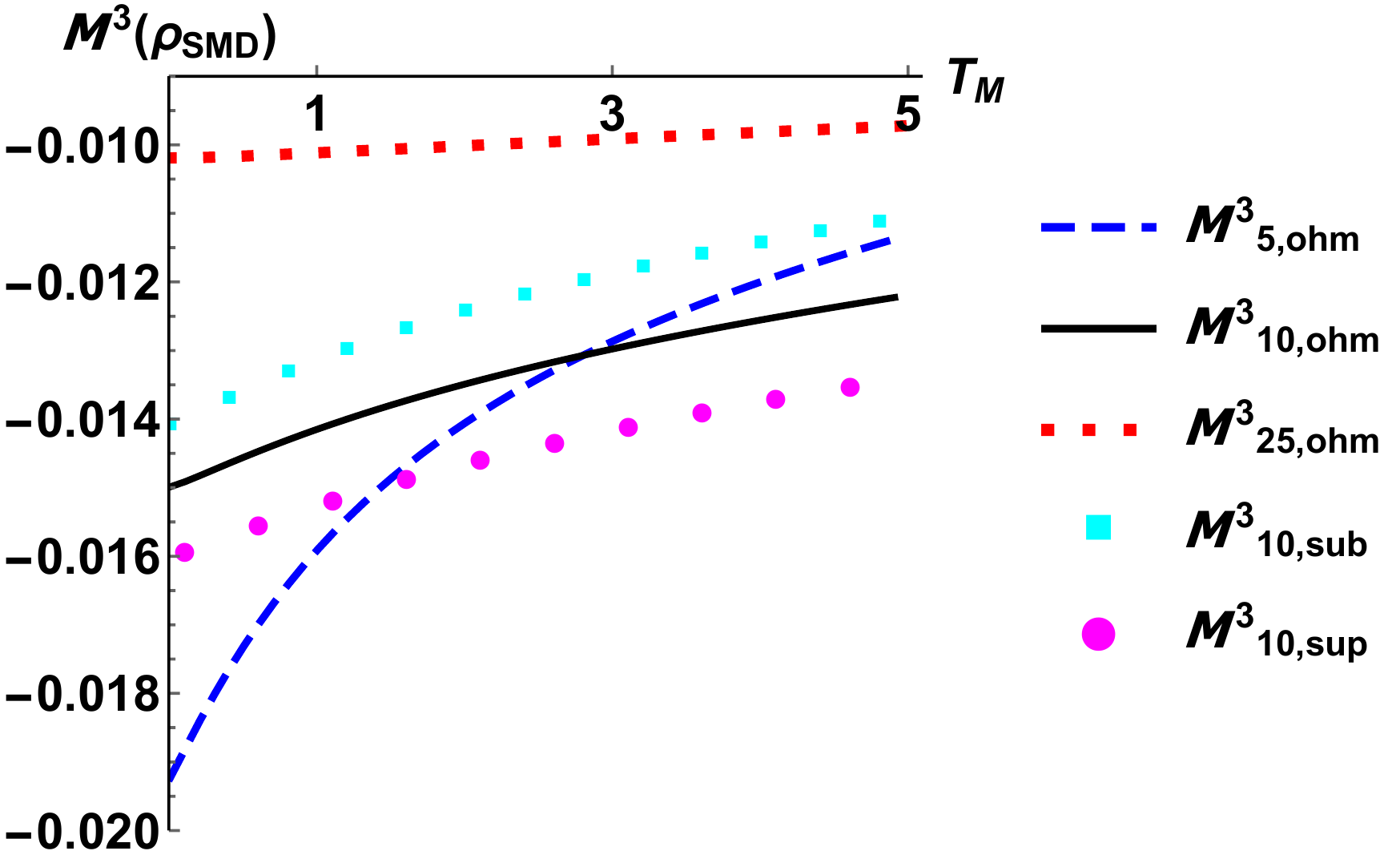}
\caption{\label{fig:mut} The tripartite mutual information defined in Equation \eqref{eq:mutualinfo3} as a function of the modulator temperature $T_M$.  All the configurations are labelled by the temperature of the source reservoir and by the type of the spectral density considered.  
}
\end{figure}

\begin{figure}[H] 
\includegraphics[width= 0.285 \columnwidth]{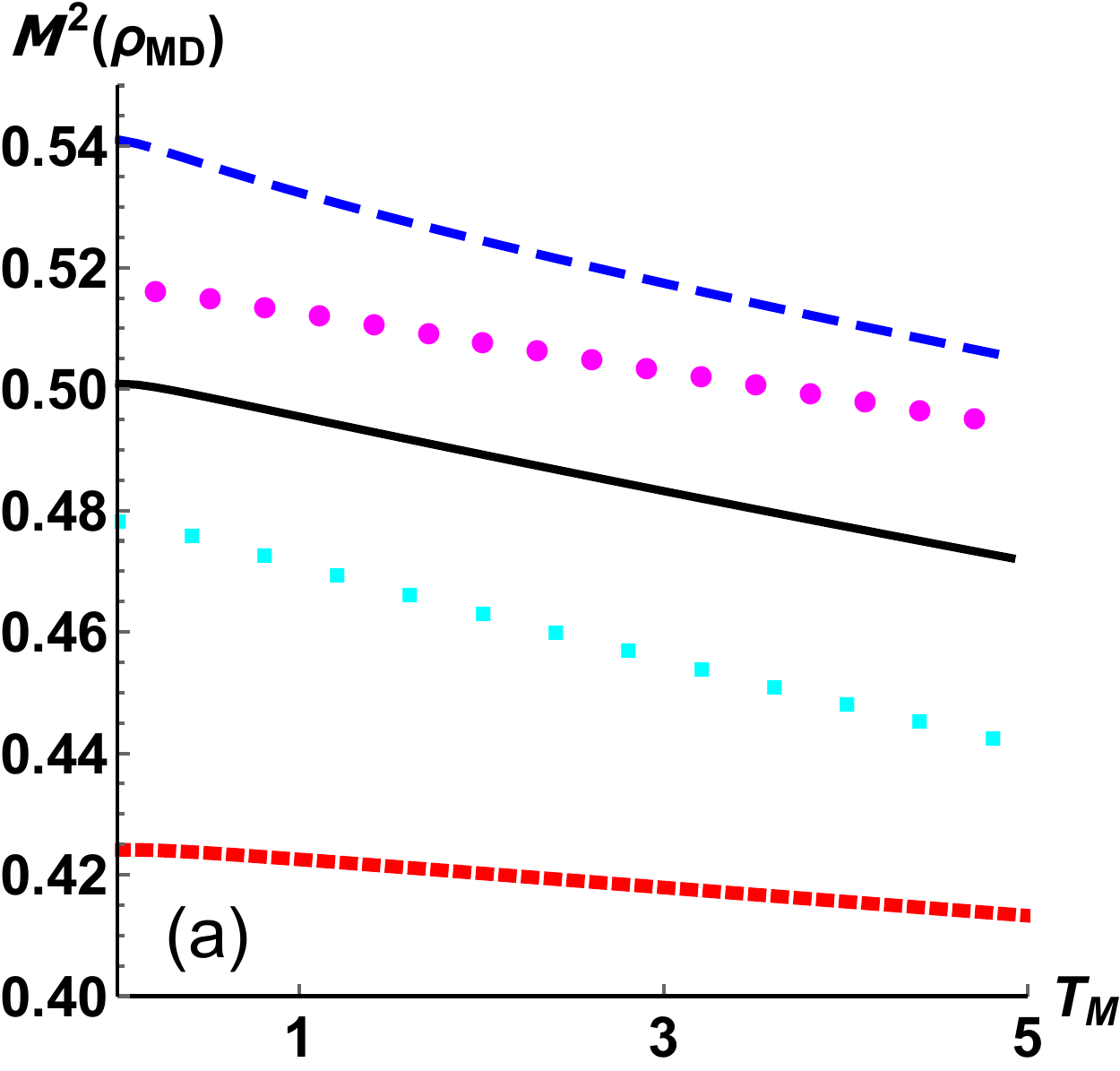}
\includegraphics[width= 0.285 \columnwidth]{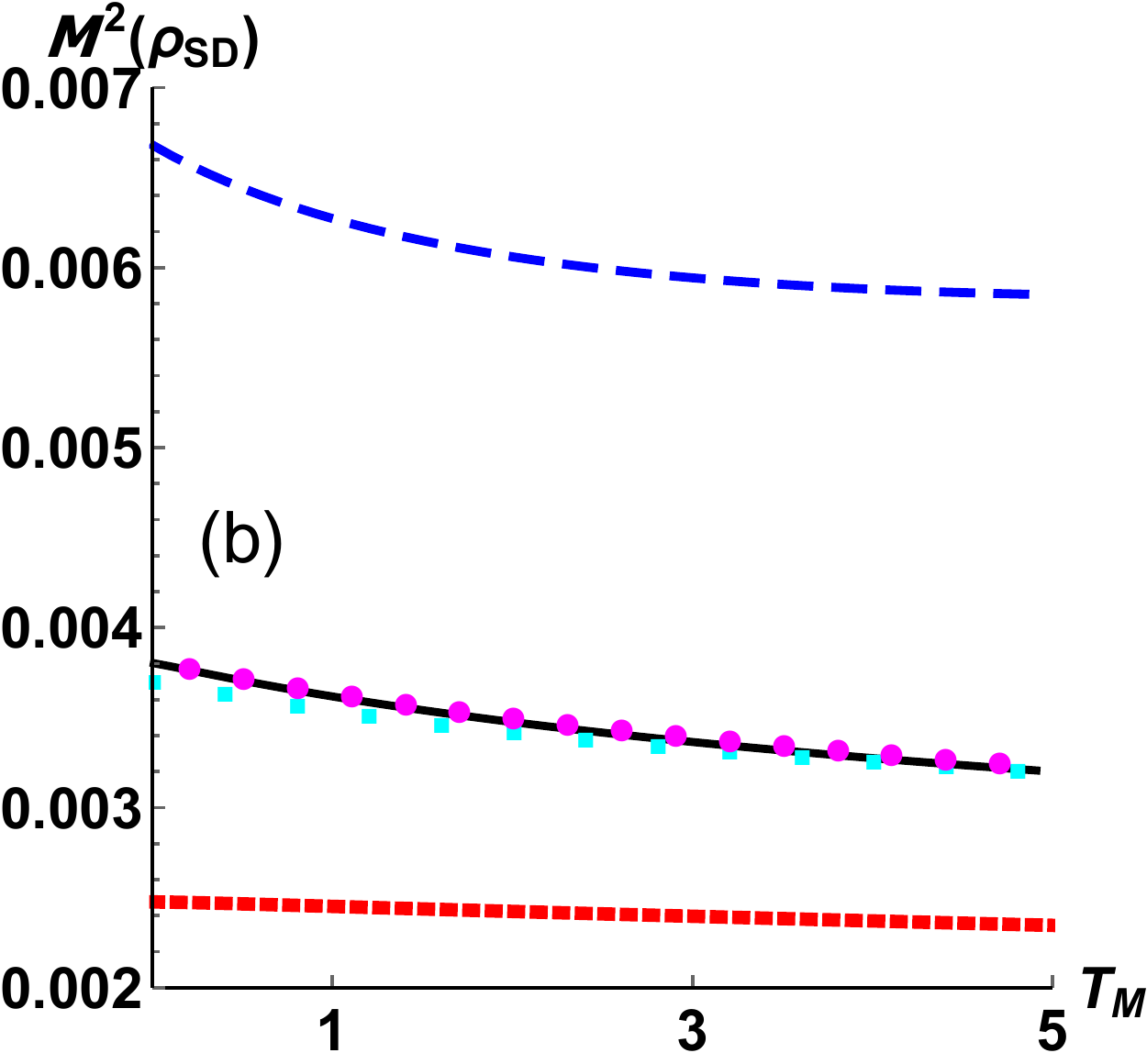}
\includegraphics[width= 0.42 \columnwidth]{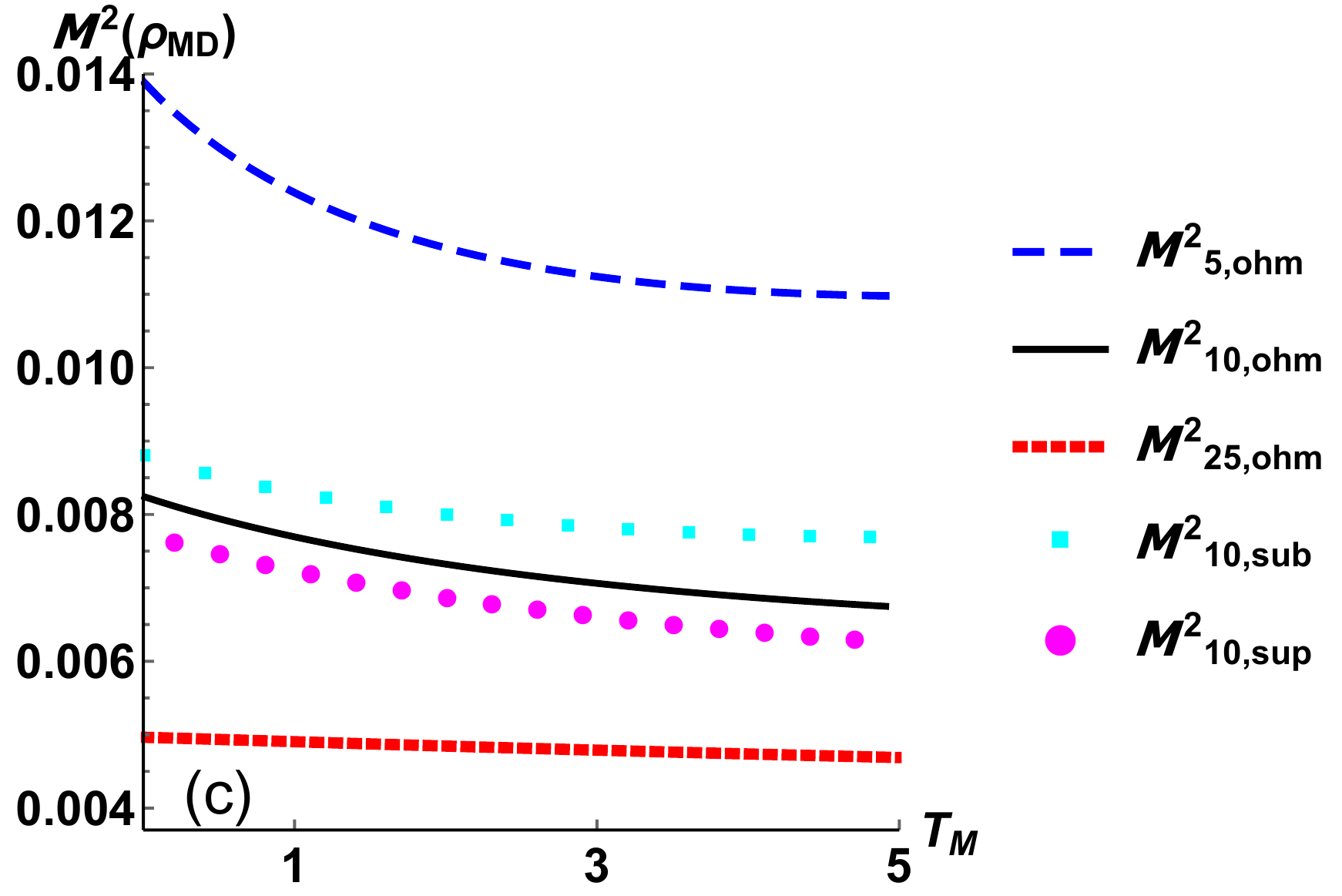}
\caption{\label{fig:mut2} The bipartite mutual information defined in Equation \eqref{eq:mutualinfo2} as a function of the modulator temperature $T_M$, when one of the three qubit is traced out. 
The plots refers to tracing out the qubit directly coupled with the source (\textbf{a}), the~modulator (\textbf{b}), the~drain (\textbf{c}).  
In the common legend the configurations are labelled by the temperature of the source reservoir and by the type of the spectral density considered.  
}
\end{figure}

We report in Figure~\ref{fig:mut} the tripartite mutual information and observe for all the configuration a negative value. 
As outlined in~\cite{mutual_arul}, this is a signature of the fact that any joint two-qubit subsystem contains more information about the third qubit than the two subsystems  individually considered. 
For our purposes, the~negative values of $\mathcal{M}_3 $ are quite strong evidences of the global and collective nature of the system relaxation by means of the interactions with three different reservoirs. 
On the other side, looking at the figures of the plots in Figure~\ref{fig:mut2} we observe that the bipartite mutual information has the same functional behaviour, 
but the subsystem composed by the modulator qubit and the drain one shares a higner value of information than the other two possible partitions. 
Overall, we observe that the correlations in a NESS stemming from an Ohmic dissipation are always sandwiched by those from the subOhmic and superOhmic model of dissipation. 
In contrast, as~intuitively expected, { a higher value of the temperature $T_S$ lowers} the total amount of~correlations. 

{The mutual information takes into account all the correlations present in the state, but~it would be also useful to have an evaluation of the purely quantum part of {them}. To~the best of our knowledge, all the measures of quantum correlations in a multipartite mixed state are mere arithmetic or geometric means of the quantum correlations in two-qubit reduced systems. For~the three-qubit state it reduces to the mean over the three possible two-qubit states obtained after a bipartition that singles out one qubit~\cite{3qubit}. Anyway, a~measure that goes beyond this approach has not been proposed yet.}

For this reason, we find more instructive to asses only the entanglement in the three bipartition, quantified via a measure based on the Peres--Horodecki criterion~\cite{peres, hhh3}. We introduce for this task the negativity~\cite{negativity2002} defined as
\begin{equation}
\label{eq:negat}
    \mathcal{N}(\rho_{AB}) = -  \sum_i |\mu_i| - \mu_i
\end{equation}
where $\mu_i$ are the eigenvalues of the matrix $\rho_{AB}^{\tau_B},$ i.e.,~the partial transpose of the matrix $\rho_{AB}$ with respect to the subsystem~B. 

\begin{figure}[H] 
\includegraphics[width= 0.285 \columnwidth]{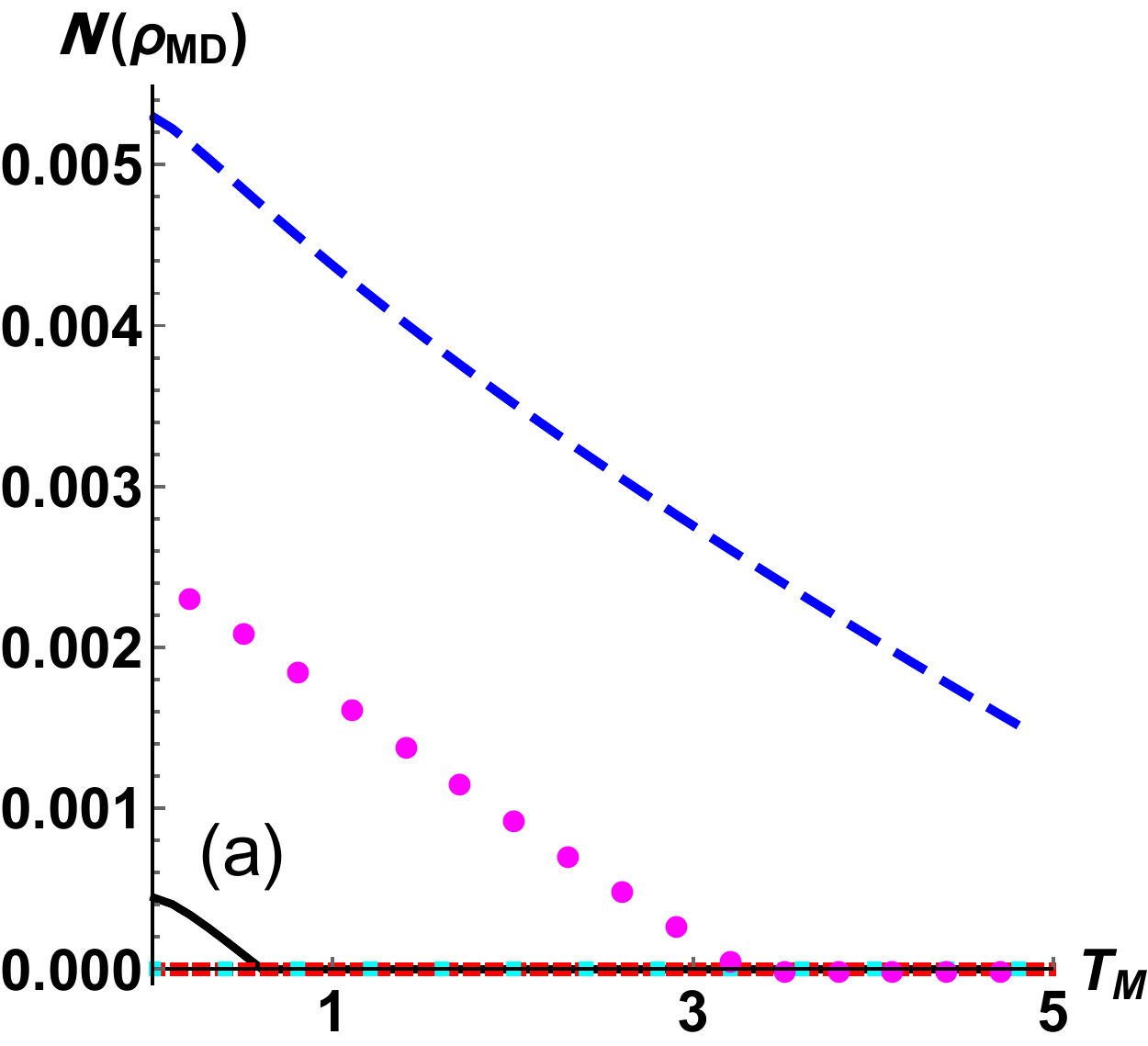}
\includegraphics[width= 0.285 \columnwidth]{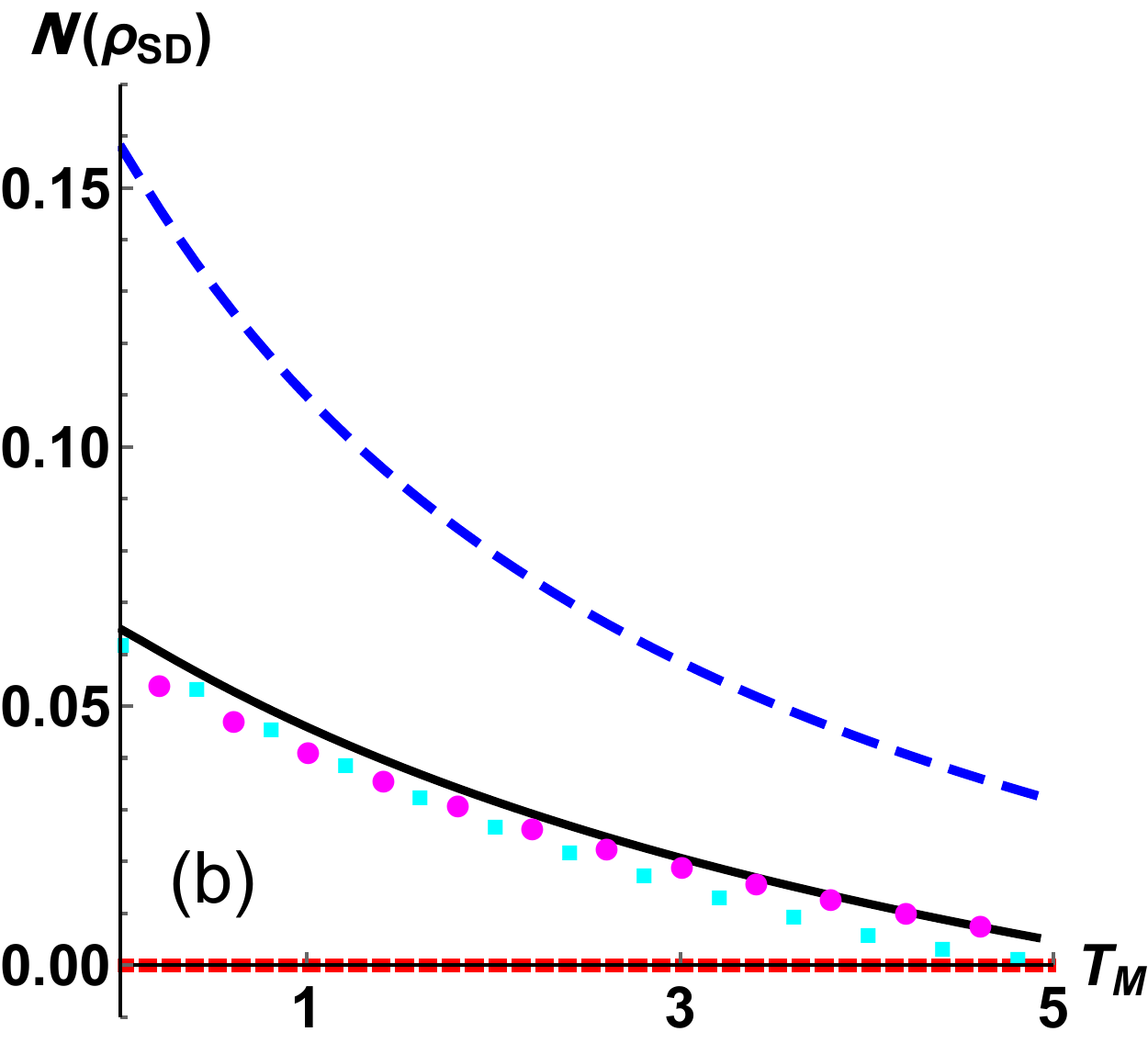}
\includegraphics[width= 0.42 \columnwidth]{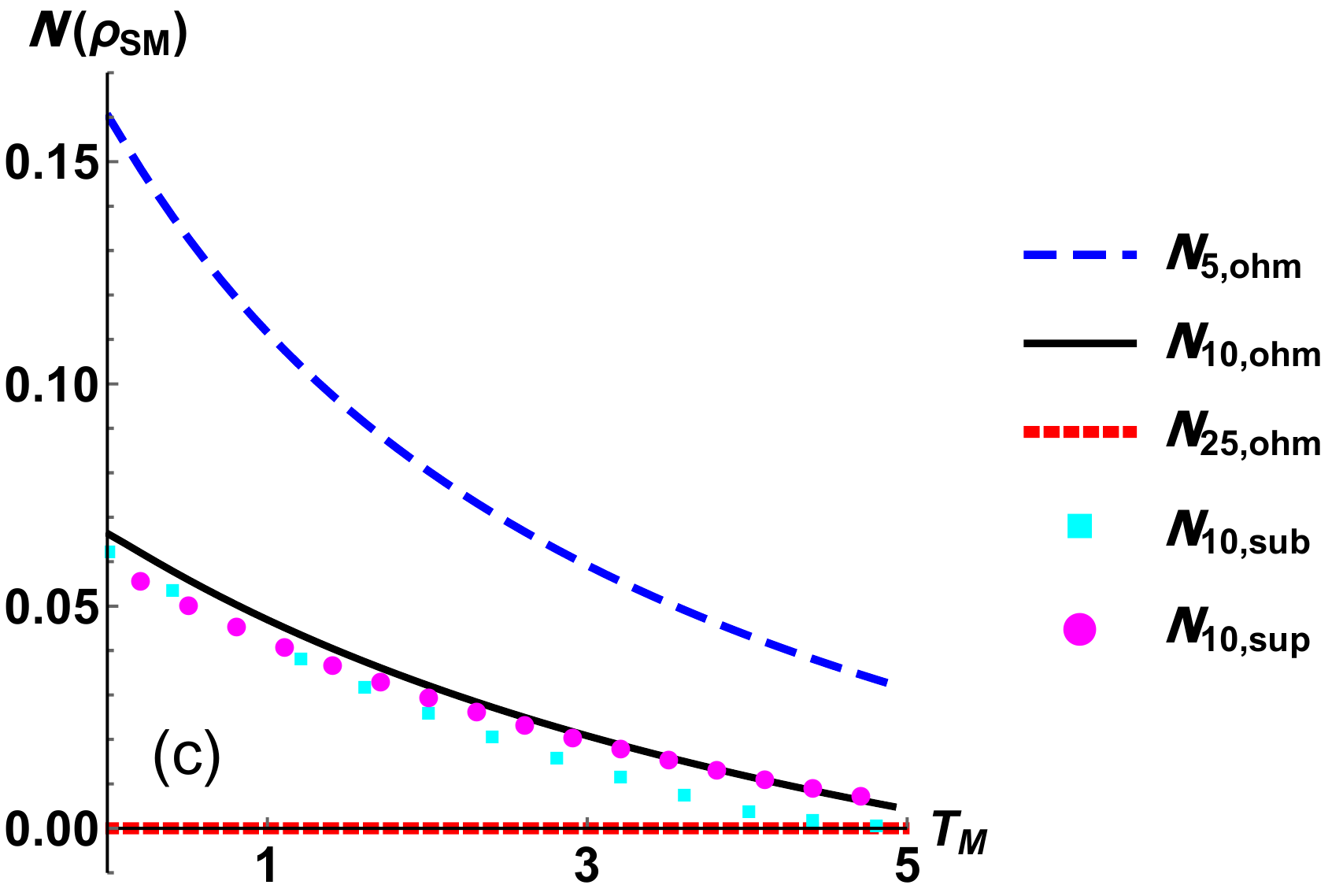}
\caption{\label{fig:neg} As measure of bipartite quantum correlation we consider the negativity defined in Equation~\eqref{eq:negat} as a function of $T_M$. The~negativity is defined for two-qubit systems so  one of the three qubits has to be traced out. 
The plots refers to the two-qubit system obtained when the qubit directly coupled with, (\textbf{a}) the source, (\textbf{b}) the modulator, (\textbf{c}) the drain is traced.   
The common legend explains the various considered settings.    
}
\end{figure}

In Figure~\ref{fig:neg} we address the quantum correlations present in the three possible partitions of the system. We observe in all the three plots that in the configuration with $T_M=25 \omega$ all the quantum correlations are almost null~\cite{Yu_2009}. Nonetheless, quantum correlations are more evident in the two subsystem with a lower value of bipartite mutual information $\mathcal{M}^2.$ 
In Appendix \ref{app:fidel}, to~better substantiate the role of the quantum correlations in the total three-qubit system, we address a straightforward comparison with two exemplary states having 
extremely non-classical~properties.

\section{Outro and Future~Perspectives}
\label{sec:Outro}
In this paper we have analyzed how different environmental settings are influencing the performance of a system behaving as QTT. The~effect is due to a purely dissipative dynamics induced by three thermal reservoirs leading the quantum system to a NESS. The~role of the dissipation has been tackled considering as main phenomenological parameters the temperature gradient between the reservoirs constituting the source and the drain, and~the spectral densities of the baths. {We have shown that, moving from a subOhmic to a superOhmic type of noise,} one induces a transition around the Ohmic regime in which the thermal transistor effect is enhanced.  Anyway, our study suggests that, at~fixed system engineering, the~best way to produce a heat current amplification is to increase the temperature gradient between the source and the {drain~reservoirs}. 

Moreover, we have observed how the correlations,  in~particular the quantum ones, among~the three subsystems do not play any fundamental role in building a quantum thermal transistor, but~on the other side they signal that the transistor effect is a collective phenomena.
We leave, as~an open problem for future investigation, the~question as to whether it is possible to engineer a system with tunable interaction, that allows us to employ a three-qubit system either as absorption refrigerator or thermal~transistor. 

As final remark, we notice that our work is an initial contribution to an interesting avenue of research constituted by bath engineering for quantum thermal analogues of electronic systems. To~boost the performances of thermal devices, one should consider a microscopic model of the reservoirs to have a better insight on the {non-equilibrium} configurations of the system, for~example in~\cite{noneq_cond}, in~terms of {non-equilibrium} Green functions to tackle {regimes} beyond the weak-coupling limit covered in the present~paper. 

\vspace{6pt}

\acknowledgments{This work is supported by Foundation for Polish Science (FNP), IRAP project ICTQT, contract no. 2018/MAB/5, co-financed by EU  Smart Growth Operational Programme and by (Polish) National Science Center (NCN): MINIATURA  DEC-2020/04/X/ST2/01794.}

\appendix
\section[\appendixname~\thesection]{}
\label{app:fidel}
\begin{figure}[H] 

\centering 
\includegraphics[width= 0.41 \columnwidth]{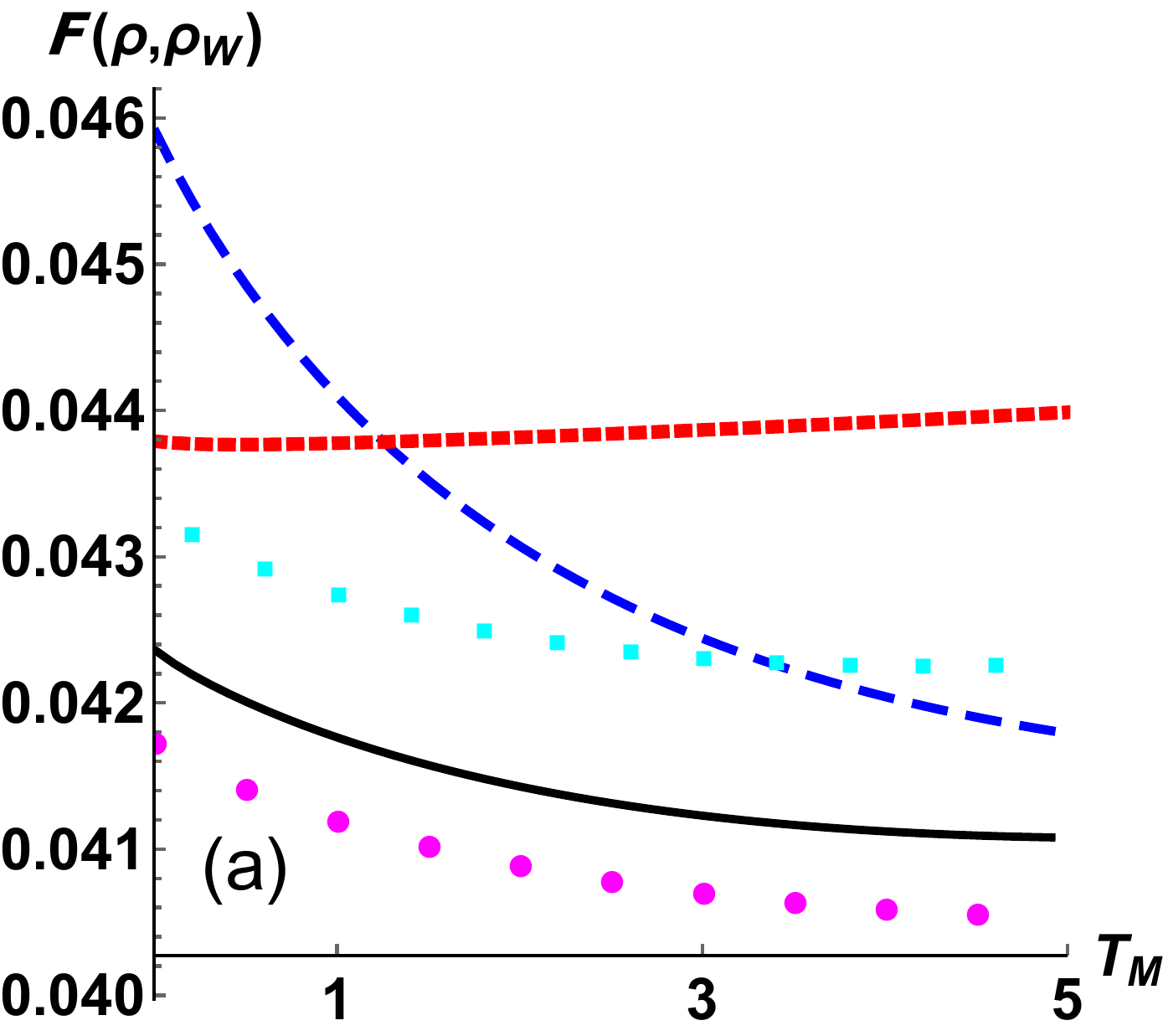}
\includegraphics[width= 0.56 \columnwidth]{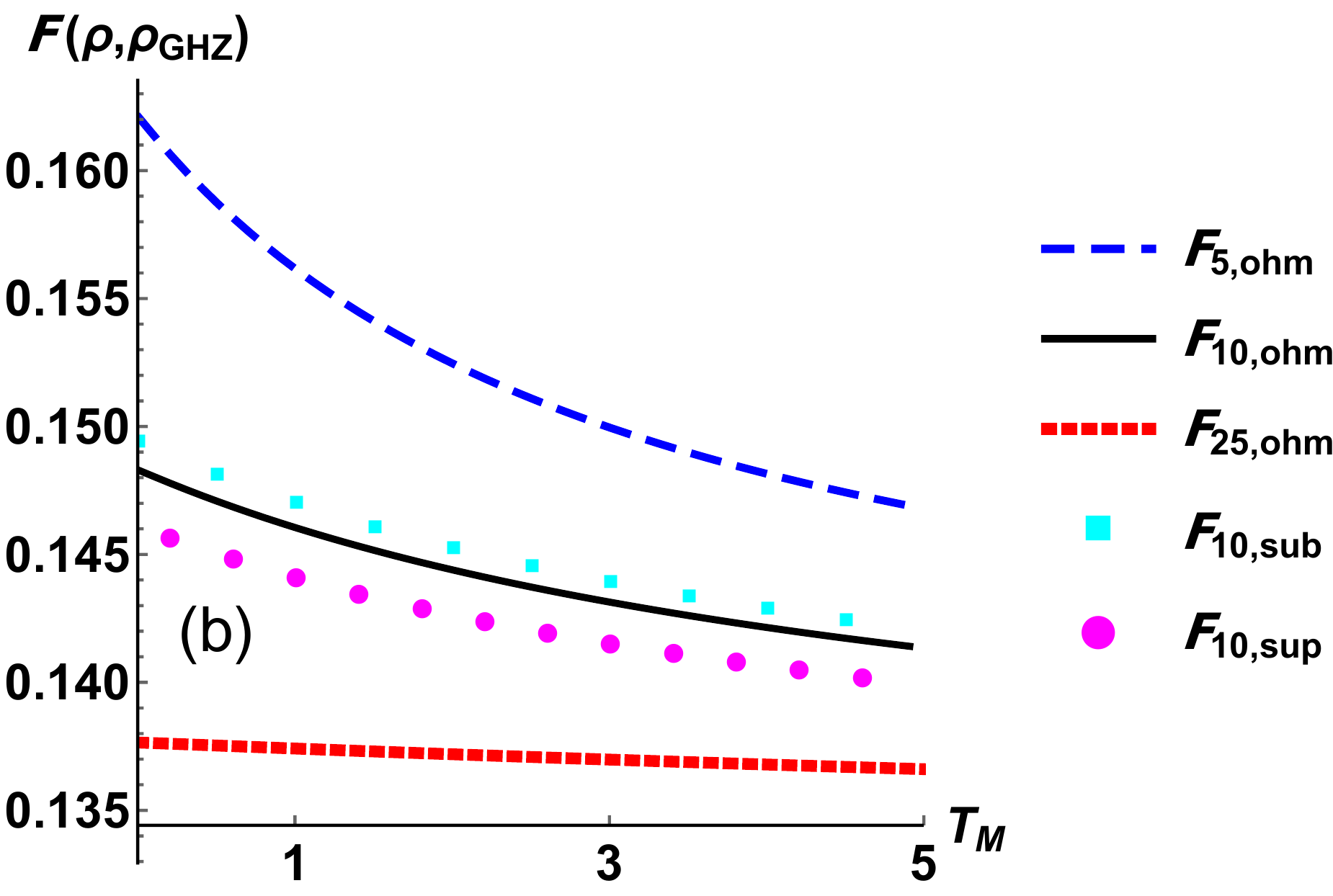}
\caption{\label{fig:fidel}The fidelity (as in Equation \eqref{eq:fid}) between the various NESS obtained when varying the modulator temperature $T_M$ and the two states representing two very different kinds of entanglement for three particles. In~(\textbf{a}) the fidelity with the $\ket{W}$ state and in (\textbf{b}) with the $\ket{GHZ}$ state. 
The common legend explains the settings we have considered in the paper.  
}
\end{figure}
The fidelity between two quantum states, $\rho$ and $\sigma,$ is defined as:
\begin{eqnarray}
\mathcal{F} = \Tr{\sqrt{\sqrt{\sigma} \rho \sqrt{\sigma}} }^2, 
\label{eq:fid}
\end{eqnarray}
and, nevertheless endowed with several limitation, the~fidelity provides a useful resource for quantum technology purposes~\cite{fid1,fid2,fid3}. 
Considering the absence of an indisputable measure of quantum correlation for mixed three qubit states, to~have a better insight on the geometry of the $\rho_{NESS}$ we compute the fidelity of the state with the the maximally nonlocal three qubits state~\cite{enriquez2018entanglement}, namely
\begin{eqnarray}
\ket{W} &=& \frac{1}{\sqrt{3}} \left[ \ket{001} + \ket{010} + \ket{100}\right]\\
\ket{GHZ}&=& \frac{1}{\sqrt{2}}\left[ \ket{000} + \ket{111}\right]. 
\end{eqnarray}

We report in Figure~\ref{fig:fidel} the fidelity between the various $\rho_{NESS}$ obtained via tuning the modulator temperature in the range $[ 0, 5]$ with the two states that are the paradigmatic example of two classes of nonclassical correlation present in a three-qubit system~\cite{cirac3q, acin3q}. 
The plots indicate that the NESS shares a higher fidelity with the $\ket{GHZ}$ state than with the $\ket{W}$. We remind that the correlation and the nonclassical properties 
of the $\ket{GHZ}$ state are, in~a certain sense, higher in respect to those of the $\ket{W}$, (the negativity of each bipartite partitions are $\mathcal{N}_{GHZ}=1$,  $\mathcal{N}_{W}=\frac{2 \sqrt{2}}{3}$,  respectively) 
but more fragile when loosing a particle. The~previous consideration allows us to surmise that this is again a signature that the scarce relevance of the the correlation present in the system on the thermal transistor effect, 
but of the relevance of the global three-particle collective dissipation in three different~baths. 





\begin{thebibliography}{999}

\bibitem[Wang and Li(2008)]{wang2008phononics}
Wang, L.; Li, B.
\newblock Phononics gets hot.
\newblock {\em Phys. World} {\bf 2008}, {\em 21},~27.

\bibitem[Li \em{et~al.}(2012)Li, Ren, Wang, Zhang, H\"anggi, and Li]{phononics}
Li, N.; Ren, J.; Wang, L.; Zhang, G.; H\"anggi, P.; Li, B.
\newblock Colloquium: Phononics: Manipulating heat flow with electronic analogs
  and beyond.
\newblock {\em Rev. Mod. Phys.} {\bf 2012}, {\em 84},~1045--1066.
\newblock {{https://doi.org/10.1103/RevModPhys.84.1045}}.

\bibitem[Dutta \em{et~al.}(2020)Dutta, Majidi, Talarico, Lo~Gullo, Courtois,
  and Winkelmann]{heatvalve}
Dutta, B.; Majidi, D.; Talarico, N.W.; Lo~Gullo, N.; Courtois, H.; Winkelmann,
  C.B.
\newblock Single-Quantum-Dot Heat Valve.
\newblock {\em Phys. Rev. Lett.} {\bf 2020}, {\em 125},~237701.
\newblock {{https://doi.org/10.1103/PhysRevLett.125.237701}}.

\bibitem[Li \em{et~al.}(2004)Li, Wang, and Casati]{diode1}
Li, B.; Wang, L.; Casati, G.
\newblock Thermal Diode: Rectification of Heat Flux.
\newblock {\em Phys. Rev. Lett.} {\bf 2004}, {\em 93},~184301.
\newblock {{https://doi.org/10.1103/ PhysRevLett.93.184301}}.

\bibitem[Pereira(2010)]{diode2}
Pereira, E.
\newblock Graded anharmonic crystals as genuine thermal diodes: Analytical
  description of rectification and negative differential thermal resistance.
\newblock {\em Phys. Rev. E} {\bf 2010}, {\em 82},~040101.
\newblock {{https://doi.org/10.1103/PhysRevE.82.040101}}.

\bibitem[Werlang \em{et~al.}(2014)Werlang, Marchiori, Cornelio, and
  Valente]{diode3}
Werlang, T.; Marchiori, M.A.; Cornelio, M.F.; Valente, D.
\newblock Optimal rectification in the ultrastrong coupling regime.
\newblock {\em Phys. Rev. E} {\bf 2014}, {\em 89},~062109.
\newblock {{https://doi.org/10.1103/PhysRevE.89.062109}}.

\bibitem[Ordonez-Miranda \em{et~al.}(2017)Ordonez-Miranda, Ezzahri, and
  Joulain]{diode4}
Ordonez-Miranda, J.; Ezzahri, Y.; Joulain, K.
\newblock Quantum thermal diode based on two interacting spinlike systems under
  different excitations.
\newblock {\em Phys. Rev. E} {\bf 2017}, {\em 95},~022128.
\newblock {{https://doi.org/10.1103/PhysRevE.95.022128}}.

\bibitem[Li \em{et~al.}(2006)Li, Wang, and Casati]{thermalTr}
Li, B.; Wang, L.; Casati, G.
\newblock Negative differential thermal resistance and thermal transistor.
\newblock {\em Appl. Phys. Lett.} {\bf 2006}, {\em 88},~143501.
\newblock {{https://doi.org/10.1063/1.2191730}}.

\bibitem[Joulain \em{et~al.}(2016)Joulain, Drevillon, Ezzahri, and
  Ordonez-Miranda]{QTT1}
Joulain, K.; Drevillon, J.; Ezzahri, Y.; Ordonez-Miranda, J.
\newblock Quantum Thermal Transistor.
\newblock {\em Phys. Rev. Lett.} {\bf 2016}, {\em 116},~200601.
\newblock {{https://doi.org/10.1103/PhysRevLett.116.200601}}.

\bibitem[Mandarino \em{et~al.}(2021)Mandarino, Joulain, G\'omez, and
  Bellomo]{QTT2}
Mandarino, A.; Joulain, K.; G\'omez, M.D.; Bellomo, B.
\newblock Thermal Transistor Effect in Quantum Systems.
\newblock {\em Phys. Rev. Appl.} {\bf 2021}, {\em 16},~034026.
\newblock {{https://doi.org/10.1103/PhysRevApplied.16.034026}}.

\bibitem[Guo \em{et~al.}(2018)Guo, Liu, and Yu]{QTT3}
Guo, B.Q.; Liu, T.; Yu, C.S.
\newblock Quantum thermal transistor based on qubit-qutrit coupling.
\newblock {\em Phys. Rev. E} {\bf 2018}, {\em 98},~022118.
\newblock {{https://doi.org/10.1103/PhysRevE.98.022118}}.

\bibitem[Majland \em{et~al.}(2020)Majland, Christensen, and Zinner]{QTT4}
Majland, M.; Christensen, K.S.; Zinner, N.T.
\newblock Quantum thermal transistor in superconducting circuits.
\newblock {\em Phys. Rev. B} {\bf 2020}, {\em 101},~184510.
\newblock {{https://doi.org/10.1103/PhysRevB.101.184510}}.

\bibitem[Wang and Li(2007)]{ThermalGate}
Wang, L.; Li, B.
\newblock Thermal Logic Gates: Computation with Phonons.
\newblock {\em Phys. Rev. Lett.} {\bf 2007}, {\em 99},~177208.
\newblock {{https://doi.org/10.1103/Phys RevLett.99.177208}}.

\bibitem[Liu \em{et~al.}(2021)Liu, Yu, and Yu]{liu2021common}
Liu, Y.Q.; Yu, D.H.; Yu, C.S.
\newblock Common environmental effects on quantum thermal transistor.
\newblock {\em Entropy} {\bf 2021}, {\em 24},~32.

\bibitem[Wijesekara \em{et~al.}(2021)Wijesekara, Gunapala, and
  Premaratne]{QTTn1}
Wijesekara, R.T.; Gunapala, S.D.; Premaratne, M.
\newblock Darlington pair of quantum thermal transistors.
\newblock {\em Phys. Rev. B} {\bf 2021}, {\em 104},~045405.
\newblock {{https://doi.org/10.1103/PhysRevB.104.045405}}.

\bibitem[Wijesekara \em{et~al.}(2022)Wijesekara, Gunapala, and
  Premaratne]{QTTn2}
Wijesekara, R.T.; Gunapala, S.D.; Premaratne, M.
\newblock Towards quantum thermal multi-transistor systems: Energy divider
  formalism.
\newblock {\em Phys. Rev. B} {\bf 2022}, {\em 105},~235412.
\newblock {{https://doi.org/10.1103/PhysRevB.105.235412}}.

\bibitem[Salvatori \em{et~al.}(2014)Salvatori, Mandarino, and Paris]{LMG14}
Salvatori, G.; Mandarino, A.; Paris, M.G.A.
\newblock Quantum metrology in Lipkin-Meshkov-Glick critical systems.
\newblock {\em Phys. Rev. A} {\bf 2014}, {\em 90},~022111.
\newblock {{https://doi.org/10.1103/PhysRevA.90.022111}}.

\bibitem[Opatrn\'y \em{et~al.}(2015)Opatrn\'y, Kol\'a\ifmmode~\check{r}\else
  \v{r}\fi{}, and Das]{appLMG1}
Opatrn\'y, T.C.V.; Kol\'a\ifmmode~\check{r}\else \v{r}\fi{}, M.; Das, K.K.
\newblock Spin squeezing by tensor twisting and Lipkin-Meshkov-Glick dynamics
  in a toroidal Bose-Einstein condensate with spatially modulated nonlinearity.
\newblock {\em Phys. Rev. A} {\bf 2015}, {\em 91},~053612.
\newblock {{https://doi.org/10.1103/PhysRevA.91. 053612}}.

\bibitem[Campbell \em{et~al.}(2015)Campbell, De~Chiara, Paternostro, Palma, and
  Fazio]{appLMG2}
Campbell, S.; De~Chiara, G.; Paternostro, M.; Palma, G.M.; Fazio, R.
\newblock Shortcut to Adiabaticity in the Lipkin-Meshkov-Glick Model.
\newblock {\em Phys. Rev. Lett.} {\bf 2015}, {\em 114},~177206.
\newblock {{https://doi.org/10.1103/PhysRevLett.114.177206}}.

\bibitem[Linden \em{et~al.}(2010)Linden, Popescu, and Skrzypczyk]{refr1}
Linden, N.; Popescu, S.; Skrzypczyk, P.
\newblock How Small Can Thermal Machines Be? The Smallest Possible
  Refrigerator.
\newblock {\em Phys. Rev. Lett.} {\bf 2010}, {\em 105},~130401.
\newblock {{https://doi.org/10.1103/PhysRevLett.105.130401}}.

{\bibitem[Brunner \em{et~al.}(2010)Huber, Linden, Popescu, Silva and Skrzypczyk]{refr2}
Brunner N.; Huber M.; Linden, N.; Silva R.; Popescu, S.; Skrzypczyk, P. \newblock Entanglement enhances cooling in microscopic quantum refrigerators.
\newblock {\em Phys. Rev. E} {\bf 2014}, {\em 89},~032115.
\newblock {{https://doi.org/10.1103/PhysRevE.89.032115}}.}

\bibitem[Correa \em{et~al.}(2014)Correa, Palao, Alonso, and Adesso]{refr3}
Correa, L.A.; Palao, J.P.; Alonso, D.; Adesso, G.
\newblock Quantum-enhanced absorption refrigerators.
\newblock {\em Sci. Rep.} {\bf 2014}, {\em 4},~3949.

\bibitem[Correa \em{et~al.}(2013)Correa, Palao, Adesso, and Alonso]{refr4}
{Correa, L.A.; Palao, J.P.; Adesso, G.; Alonso, D.}
\newblock Performance bound for quantum absorption refrigerators.
\newblock {\em Phys. Rev. E} {\bf 2013}, {\em 87},~042131.
\newblock {{https://doi.org/10.1103/PhysRevE.87.042131}}.

\bibitem[Correa \em{et~al.}(2014)Correa, Palao, Adesso, and Alonso]{refr5}
Correa, L.A.; Palao, J.P.; Adesso, G.; Alonso, D.
\newblock Optimal performance of endoreversible quantum refrigerators.
\newblock {\em Phys. Rev. E} {\bf 2014}, {\em 90},~062124.
\newblock {{https://doi.org/10.1103/PhysRevE.90.062124}}.

\bibitem[Galve \em{et~al.}(2017)Galve, Mandarino, Paris, Benedetti, and
  Zambrini]{galve2017}
Galve, F.; Mandarino, A.; Paris, M.G.; Benedetti, C.; Zambrini, R.
\newblock Microscopic description for the emergence of collective dissipation
  in extended quantum systems.
\newblock {\em Sci. Rep.} {\bf 2017}, {\em 7},~42050.

\bibitem[Davies(1974)]{davies1974markovian}
Davies, E.B.
\newblock Markovian master equations.
\newblock {\em Comm. Math. Phys.} {\bf 1974}, {\em 39},~91--110.

\bibitem[Alicki and Lendi(2007)]{alicki2007quantum}
Alicki, R.; Lendi, K.
\newblock {\em Quantum Dynamical Semigroups and Applications}; Springer: {Berlin/Heidelberg, Germany,}
  2007; Volume 717.

\bibitem[Gorini \em{et~al.}(1976)Gorini, Kossakowski, and Sudarshan]{GKSL1}
Gorini, V.; Kossakowski, A.; Sudarshan, E.C.G.
\newblock Completely positive dynamical semigroups of N-level systems.
\newblock {\em J. Math. Phys.} {\bf 1976}, {\em 17},~821--825.

\bibitem[Lindblad(1976)]{GKSL2}
Lindblad, G.
\newblock On the generators of quantum dynamical semigroups.
\newblock {\em Commun. Math. Phys.} {\bf 1976}, {\em 48},~119--130.

\bibitem[Leggett \em{et~al.}(1987)Leggett, Chakravarty, Dorsey, Fisher, Garg,
  and Zwerger]{Rev_Leggett}
Leggett, A.J.; Chakravarty, S.; Dorsey, A.T.; Fisher, M.P.A.; Garg, A.;
  Zwerger, W.
\newblock Dynamics of the dissipative two-state system.
\newblock {\em Rev. Mod. Phys.} {\bf 1987}, {\em 59},~1--85.
\newblock {{https://doi.org/10.1103/RevModPhys.59.1}}.

\bibitem[Caldeira and Leggett(1981)]{caldeira1981}
Caldeira, A.O.; Leggett, A.J.
\newblock Influence of Dissipation on Quantum Tunneling in Macroscopic Systems.
\newblock {\em Phys. Rev. Lett.} {\bf 1981}, {\em 46},~211--214.
\newblock {{https://doi.org/10.1103/PhysRevLett.46.211}}.

\bibitem[Caldeira and Leggett(1983)]{caldeira1983}
Caldeira, A.O.; Leggett, A.J.
\newblock Quantum tunnelling in a dissipative system.
\newblock {\em Ann. Phys.} {\bf 1983}, {\em 149},~374--456.

\bibitem[Wilner \em{et~al.}(2015)Wilner, Wang, Thoss, and Rabani]{SD1}
Wilner, E.Y.; Wang, H.; Thoss, M.; Rabani, E.
\newblock Sub-Ohmic to super-Ohmic crossover behavior in nonequilibrium quantum
  systems with electron-phonon interactions.
\newblock {\em Phys. Rev. B} {\bf 2015}, {\em 92},~195143.
\newblock {{https://doi.org/10.1103/PhysRevB.92.195143}}.

\bibitem[Weiss(2012)]{Weiss}
Weiss, U.
\newblock {\em Quantum Dissipative Systems}, 4th ed.; World Scientific: {Singapore}, 2012.
\newblock {{https://doi.org/10.1142/8334}}.

\bibitem[Kosloff(2013)]{kosloff}
Kosloff, R.
\newblock Quantum thermodynamics: A dynamical viewpoint.
\newblock {\em Entropy} {\bf 2013}, {\em 15},~2100--2128.

\bibitem[Rivas(2019)]{rivas2019}
Rivas, {\'A}.
\newblock Quantum thermodynamics in the refined weak coupling limit.
\newblock {\em Entropy} {\bf 2019}, {\em 21},~725.

\bibitem[Seshadri \em{et~al.}(2018)Seshadri, Madhok, and
  Lakshminarayan]{mutual_arul}
Seshadri, A.; Madhok, V.; Lakshminarayan, A.
\newblock Tripartite mutual information, entanglement, and scrambling in
  permutation symmetric systems with an application to quantum chaos.
\newblock {\em Phys. Rev. E} {\bf 2018}, {\em 98},~052205.
\newblock {{https://doi.org/10.1103/PhysRevE.98.052205}}.

\bibitem[Rota(2016)]{rota2016tripartite}
Rota, M.
\newblock Tripartite information of highly entangled states.
\newblock {\em J. High Energy Phys.} {\bf 2016}, {\em 2016},~75.

\bibitem[Rangamani and Rota(2015)]{Rangamani_2015}
Rangamani, M.; Rota, M.
\newblock Entanglement structures in qubit systems.
\newblock {\em J. Phys. A Math. Theor.} {\bf 2015},
  {\em 48},~385301.
\newblock {{https://doi.org/10.1088/1751-8113/48/38/385301}}.

\bibitem[Kalaga \em{et~al.}(2022)Kalaga, Leo{\'n}ski, Szcz\k{e}{\'s}niak, and
  Pe{\v{r}}ina~Jr]{3qubit}
Kalaga, J.K.; Leo{\'n}ski, W.; Szcz\k{e}{\'s}niak, R.; Pe{\v{r}}ina~Jr, J.
\newblock Mixedness, Coherence and Entanglement in a Family of Three-Qubit
  States.
\newblock {\em Entropy} {\bf 2022}, {\em 24},~324.

\bibitem[Peres(1996)]{peres}
Peres, A.
\newblock Separability Criterion for Density Matrices.
\newblock {\em Phys. Rev. Lett.} {\bf 1996}, {\em 77},~1413--1415.
\newblock {{https://doi.org/10.1103/PhysRevLett.77. 1413}}.

\bibitem[Horodecki \em{et~al.}(1996)Horodecki, Horodecki, and Horodecki]{hhh3}
Horodecki, M.; Horodecki, P.; Horodecki, R.
\newblock Separability of mixed states: Necessary and sufficient conditions.
\newblock {\em Phys. Lett. A} {\bf 1996}, {\em 223},~1--8.
\newblock
  {{https://doi.org/https://doi.org/10.1016/S0375-9601(96)00706-2}}.

\bibitem[Vidal and Werner(2002)]{negativity2002}
Vidal, G.; Werner, R.F.
\newblock Computable measure of entanglement.
\newblock {\em Phys. Rev. A} {\bf 2002}, {\em 65},~032314.
\newblock {{https://doi.org/10.1103/PhysRevA. 65.032314}}.

\bibitem[Yu and Eberly(2009)]{Yu_2009}
Yu, T.; Eberly, J.H.
\newblock Sudden Death of Entanglement.
\newblock {\em Science} {\bf 2009}, {\em 323},~598--601.
\newblock {{https://doi.org/10.1126/science.1167343}}.

\bibitem[Wang \em{et~al.}(2006)Wang, Wang, and Zeng]{noneq_cond}
Wang, J.S.; Wang, J.; Zeng, N.
\newblock Nonequilibrium Green's function approach to mesoscopic thermal
  transport.
\newblock {\em Phys. Rev. B} {\bf 2006}, {\em 74},~033408.
\newblock {{https://doi.org/10.1103/PhysRevB.74.033408}}.

\bibitem[Bina \em{et~al.}(2014)Bina, Mandarino, Olivares, and Paris]{fid1}
Bina, M.; Mandarino, A.; Olivares, S.; Paris, M.G.A.
\newblock Drawbacks of the use of fidelity to assess quantum resources.
\newblock {\em Phys. Rev. A} {\bf 2014}, {\em 89},~012305.
\newblock {{https://doi.org/10.1103/PhysRevA.89.012305}}.

\bibitem[Mandarino \em{et~al.}(2014)Mandarino, Bina, Olivares, and Paris]{fid2}
Mandarino, A.; Bina, M.; Olivares, S.; Paris, M.G.
\newblock About the use of fidelity in continuous variable systems.
\newblock {\em Int. J. Quantum Inf.} {\bf 2014}, {\em
  12},~1461015.

\bibitem[Mandarino \em{et~al.}(2016)Mandarino, Bina, Porto, Cialdi, Olivares,
  and Paris]{fid3}
Mandarino, A.; Bina, M.; Porto, C.; Cialdi, S.; Olivares, S.; Paris, M.G.A.
\newblock Assessing the significance of fidelity as a figure of merit in
  quantum state reconstruction of discrete and continuous-variable systems.
\newblock {\em Phys. Rev. A} {\bf 2016}, {\em 93},~062118.
\newblock {{https://doi.org/10.1103/PhysRevA.93.062118}}.

\bibitem[Enr{\'\i}quez \em{et~al.}(2018)Enr{\'\i}quez, Delgado, and
  {\.Z}yczkowski]{enriquez2018entanglement}
Enr{\'\i}quez, M.; Delgado, F.; {\.Z}yczkowski, K.
\newblock Entanglement of three-qubit random pure states.
\newblock {\em Entropy} {\bf 2018}, {\em 20},~745.

\bibitem[D\"ur \em{et~al.}(2000)D\"ur, Vidal, and Cirac]{cirac3q}
D\"ur, W.; Vidal, G.; Cirac, J.I.
\newblock Three qubits can be entangled in two inequivalent ways.
\newblock {\em Phys. Rev. A} {\bf 2000}, {\em 62},~062314.
\newblock {{https://doi.org/10.1103/PhysRevA.62.062314}}.

\bibitem[Ac\'{\i}n \em{et~al.}(2001)Ac\'{\i}n, Bru\ss{}, Lewenstein, and
  Sanpera]{acin3q}
Ac\'{\i}n, A.; Bru\ss{}, D.; Lewenstein, M.; Sanpera, A.
\newblock Classification of Mixed Three-Qubit States.
\newblock {\em Phys. Rev. Lett.} {\bf 2001}, {\em 87},~040401.
\newblock {{https://doi.org/10.1103/PhysRevLett.87.040401}}.

\end{thebibliography}
\end{document}